\documentclass[twocol]{ametsoc}
\newcommand*{\rom}[1]{\expandafter\@slowromancap\romannumeral #1@}
\makeatother

\usepackage{physics}
\usepackage{xspace}
\bibpunct{(}{)}{;}{a}{}{,}
\journal{jas}
\title{Can Existing Theory Predict the Response of Tropical Cyclone Intensity to Idealized Landfall?}

    \authors{Jie Chen\correspondingauthor{Jie Chen, 
     Purdue University, 
     550 Stadium Mall Dr., West Lafayette, IN 47907.}
 and Daniel R. Chavas}

     \affiliation{Purdue University, 
     West Lafayette, Indiana}

\email{chen2340@purdue.edu}

\abstract{Tropical cyclones cause significant inland hazards, including wind damage and freshwater flooding, that depend strongly on how storm intensity evolves at and after landfall. Existing theoretical predictions for the time-dependent and equilibrium response of storm intensity have been tested over the open ocean but not yet to be applied to storms after landfall. Recent work examined the transient response of the tropical cyclone low-level wind field to instantaneous surface roughening or drying in idealized axisymmetric $f$-plane simulations. Here, experiments testing combined surface roughening and drying with varying magnitudes of each are used to test theoretical predictions for the intensity response. The transient response to combined surface forcings can be reproduced by the product of their individual responses, in line with traditional potential intensity theory. Existing intensification theory is generalized to weakening and found capable of reproducing the time-dependent inland intensity decay. The initial (0-10min) rapid decay of near-surface wind caused by surface roughening is not captured by existing theory but can be reproduced by a simple frictional spin-down model, where the decay rate is a function of surface roughness. Finally, the theory is shown to compare well with the prevailing empirical decay model for real-world storms. Overall, results indicate the potential for existing theory to predict how tropical cyclone intensity evolves after landfall.} 

\begin{document}

\maketitle

%
\section{Introduction}

Landfalling tropical cyclones (TCs) bring tremendous damage to both coastal and inland regions \citep{Rappaport2000,Rappaport2014,Villarini2014}. These damages may change in the future should TCs move and/or decay more slowly in a warming climate \citep{Kossin2018,Kossin2019, Li2020}. Therefore, a credible estimation of TC intensity decay after landfall is essential for hazard prediction. Having a physically-based theoretical solution for the storm intensity response to landfall could help improve risk assessment, both in real time for impending landfall events and in climatological studies \citep{Jing2020,Xi2020}. However, the underlying physics governing this post-landfall response are not well understood, and no such predictive theory currently exists. 

Past research has examined TC intensity at and after landfall via numerical simulations of historical cases and statistical models. Numerical models have limited capacity to predict post-landfall intensity due to the difficulty of capturing the physics over complex terrain, which is not necessarily improved by increasing model resolution or assimilating observational data \citep{Shen2005,Liu1997}. Moreover, site-specific case studies are not readily generalized to a more fundamental understanding of the storm response to landfall over a wide range of land surfaces. An alternative approach is the use of empirical models or probabilistic models to predict storm inland intensity decay, which may incorporate both storm and environmental parameters \citep{Kaplan+DeMaria1995,Kaplan+DeMaria2001, Vickery1995,Vickery2005,DeMaria2006,Bhowmik2005}. Empirical models models have been incorporated into the Statistical Hurricane Intensity Prediction Scheme for the Atlantic and eastern Pacific Oceans \citep{DeMaria1994,DeMaria2005} and Statistical Typhoon Intensity Prediction Scheme for the western North Pacific \citep{Knaff2005}, and have been tested for landfalling hurricanes along the South China Coast \citep{Wong2008}. Statistical models have also been incorporated into hurricane risk assessment models in the context of climate change \citep{Vickery2000, Emanuel2006, Jing2020}. However, current statistical models do not incorporate the physics of hurricane intensity decay over land, and their accuracy is limited by the data collected to train the model. Thus, empirical models offer limited fundamental understanding of the inland decay of hurricane intensity, particularly under a changing climate. 

Physically-based theoretical models are formulated for TCs over the ocean. Quasi-steady state theories for the tropical cyclone date back to \citeauthor{Lilly1985} (1985, unpublished manuscript), \cite{Shutts1981}, \citeauthor{Emanuel1986} (1986, hereafter referred to E86). More recently, theory now exists for TC intensification over the ocean (\cite{Emanuel2012}, hereafter referred to E12) that was found to compare well with an axisymmetric model simulation \citep{Emanuel2018}. However, due to the complexities in the transition from ocean to land, research has yet to develop a theory for post-landfall decay that account for the basic physics of the response of a tropical cyclone to landfall. Meanwhile, the potential for existing theoretical models for storms over the ocean to be applied after landfall has yet to be explored. Therefore, testing existing theories against idealized landfalls is a natural step to understand how known physics can or cannot explain the response of TC intensity after landfall. This is the focus of our work.

\citeauthor{Chen+Chavas2020} (2020, hereafter referred to CC20) idealized landfall as a transient response of a mature axisymmetric TC to instantaneous surface forcing: surface roughening or drying, each over a range of magnitudes. They tested the response to each forcing individually and showed that each ultimately causes the storm to weaken but via different mechanistic pathways. They further showed that the final equilibrium intensity in response to each forcing can be predicted by E86 potential intensity theory. A logical next step is to test whether the transient intensity response can be predicted by existing theory and, further, whether results can be generalized to any combination of surface drying and roughening adjusted simultaneously. Both outcomes would be more directly relevant to a wide range of inland surfaces felt by storms in real-world landfalls.  

Therefore, in this work, both steady-state intensity theory (E86) and time-dependent intensity change theory (E12) are tested against different sets of simulations where surface roughness and wetness are individually/simultaneously modified instantaneously beneath a mature axisymmetric tropical cyclone. We seek to answer the following research questions:
\begin{enumerate}
    \item Can traditional potential intensity theory predict the equilibrium response to simultaneous surface drying and roughening?
    \item Can the transient response to simultaneous drying and roughening be predicted from the responses to each forcing individually?
    \item Can existing intensification theory predict the transient decay response to surface drying and/or roughening?
    \item Do the theories work for the intensity both near the surface and near the top of the boundary layer?
\end{enumerate}

This paper is structured as follows. Section \ref{existing solutions} reviews the relevant theories and demonstrates how they may be applied to predict the intensity response to surface forcings. Section \ref{method} describes our idealized simulation experiments that are used to test the theory. Section \ref{results} presents our results addressing the research questions. Section \ref{summary} summarizes key results, limitations, and avenues for future work.


\section{Theory}\label{existing solutions}

This work examines two existing theories that predict the equilibrium intensity (E86) and the time-dependent intensity change (E12) of a tropical cyclone. The original motivation of such theories is for storms over the ocean. CC20 found that the equilibrium response of a mature TC to instantaneous surface roughening or drying followed the response predicted by E86 theory closely. This work expands on CC20 by testing both the E86 and E12 theory and generalizes the experiments to simultaneous surface roughening and drying. This section reviews each theoretical prediction and demonstrates how they can be formulated to apply to idealized landfall experiments. 

\subsection{Equilibrium intensity prediction: E86}

Potential intensity is a theoretical upper-bound for the tropical cyclone intensity in a given thermodynamic environment. This theory is formulated by idealizing a mature tropical cyclone as a Carnot heat engine, where entropy fluxes from the ocean surface are used to maintain the circulation against surface frictional dissipation. Potential intensity $V_p$ is expressed as \citep{Bister1998}, 
\begin{equation}\label{Eq-PI}
    V_p=\sqrt{\frac{C_k}{C_d}\eta(\triangle {k})} 
\end{equation}
where
\begin{equation}\label{EQ_eta}
\eta=\frac{T_{ST}-T_{tpp}}{T_{tpp}}
\end{equation}
\begin{equation}\label{EQvp_k}
    \triangle{k}={C_p}(T_{ST}-T_a)+ \epsilon L_v(q^*(T_{ST})-q_a(T_a))
\end{equation}
$C_k$ and $C_d$ are bulk exchange coefficient for surface enthalpy and
momentum, respectively; $\triangle{k}$ is the difference between the saturation enthalpy of the ocean surface and the enthalpy of the overlying near-surface air; $T_{ST}$ is the surface temperature, $T_a$ is the temperature of air overlying the surface; $T_{tpp}$ is the tropopause temperature; $L_v$ is the enthalpy of vaporization; $C_p$ is the specific heat capacity of air; $q^*$ is the saturation mixing ratio of the ocean surface at the local surface pressure; $q_a$ is the mixing ratio of air overlying the ocean surface; and $\epsilon$ is surface evaporative fraction, which is taken as $1$ over ocean surface. This formulation includes the effect of dissipative heating, which results in the tropopause temperature replacing the surface temperature in the denominator of the Carnot efficiency factor \citep{Bister1998}.

For inland storms, where the underlying surface has a higher $C_d$ (i.e., a rougher surface) and smaller $\epsilon$ (i.e., a drier surface), Eq.\eqref{Eq-PI} would predict a weaker maximum sustained wind speed compared to the ocean environment. Thus, we define the response of $V_p$ to any given surface forcing as the ratio of its post-forcing value to its initial pre-forcing value. Following CC20, the predicted $V_p$ response to surface roughening is given by  
\begin{equation}\label{Vp_separate2}
   \tilde V_{p(C_d)}=\frac{V_{p(C_d)}}{V_{p(CTRL)}}=\sqrt \frac{C_d}{(C_d)_{EXP}}
\end{equation}
or surface drying given by
\begin{equation}\label{Vp_separate1}
   \tilde V_{p(\epsilon)}=\frac{V_{p(\epsilon)}}{V_{p(CTRL)}}=\sqrt{\frac{C_p\triangle T + (\epsilon)_{EXP} L_v\triangle{q}}{\triangle k}}
\end{equation}
where $(C_d)_{EXP}$ and $(\epsilon)_{EXP}$ are the modified value of each respective surface property parameter, representing various magnitudes of surface roughening or drying. $V_{p(CTRL)}$ is the potential intensity of a Control pre-landfall TC (defined in Section \ref{method}). As discussed in CC20, using normalized responses of $V_p$ can generalize the results to any mature storm intensity and also minimizes sensitivities associated with the precise definition of $V_p$. CC20 demonstrated that the final equilibrium intensity response in idealized simulations closely follows this prediction.

Here we expand this approach to simultaneous surface drying and roughening, i.e.
\begin{equation}\label{predictedVp}
   \tilde V_{p(C_d\epsilon)}=\frac{{V_{p(C_d\epsilon)}}}{{V_p}_{CTRL}}=\frac{\sqrt{\frac{C_k}{(C_d)_{EXP}}\eta (C_p\triangle T + (\epsilon)_{EXP} L_v\triangle{q})}}{\sqrt{\frac{C_k}{C_d}\eta \triangle k}}
\end{equation}
Mathematically, Eq.\eqref{predictedVp} indicates that the equilibrium intensity response to simultaneous forcing, $\tilde V_{p(C_d\epsilon)}$, is simply the product of the individual equilibrium responses, $\tilde V_{p(C_d)}$ and $\tilde V_{p(\epsilon)}$, i.e.
\begin{equation}\label{Vp_deconstruct}
   \tilde V_{p(C_d\epsilon)}= \tilde V_{p(C_d)}\tilde V_{p(\epsilon)}.
\end{equation}  
The implication is that the response to any combination of surface roughening and drying may be predictable from the responses to each individual forcing. Inspired by Eq.\eqref{Vp_deconstruct}, the complete time-dependent response of storm intensity $\tilde v_{m(C_d\epsilon)}(\tau)$ to simultaneous surface roughening and drying is hypothesized as the product of the individual transient responses, i.e.
\begin{equation}\label{vm_deconstruct}
 \tilde v_{m(C_d\epsilon)}(\tau)\approx \tilde v^{*}(\tau)=\tilde v_{m(C_d)}(\tau)\tilde v_{m(\epsilon)}(\tau)
\end{equation}
The above outcome would have significant practical benefits for understanding the response to a wide range of surface properties as is found in nature.

\subsection{Transient intensity prediction: E12}

\begin{figure}[t]
\centerline{\includegraphics[width=0.45\textwidth]{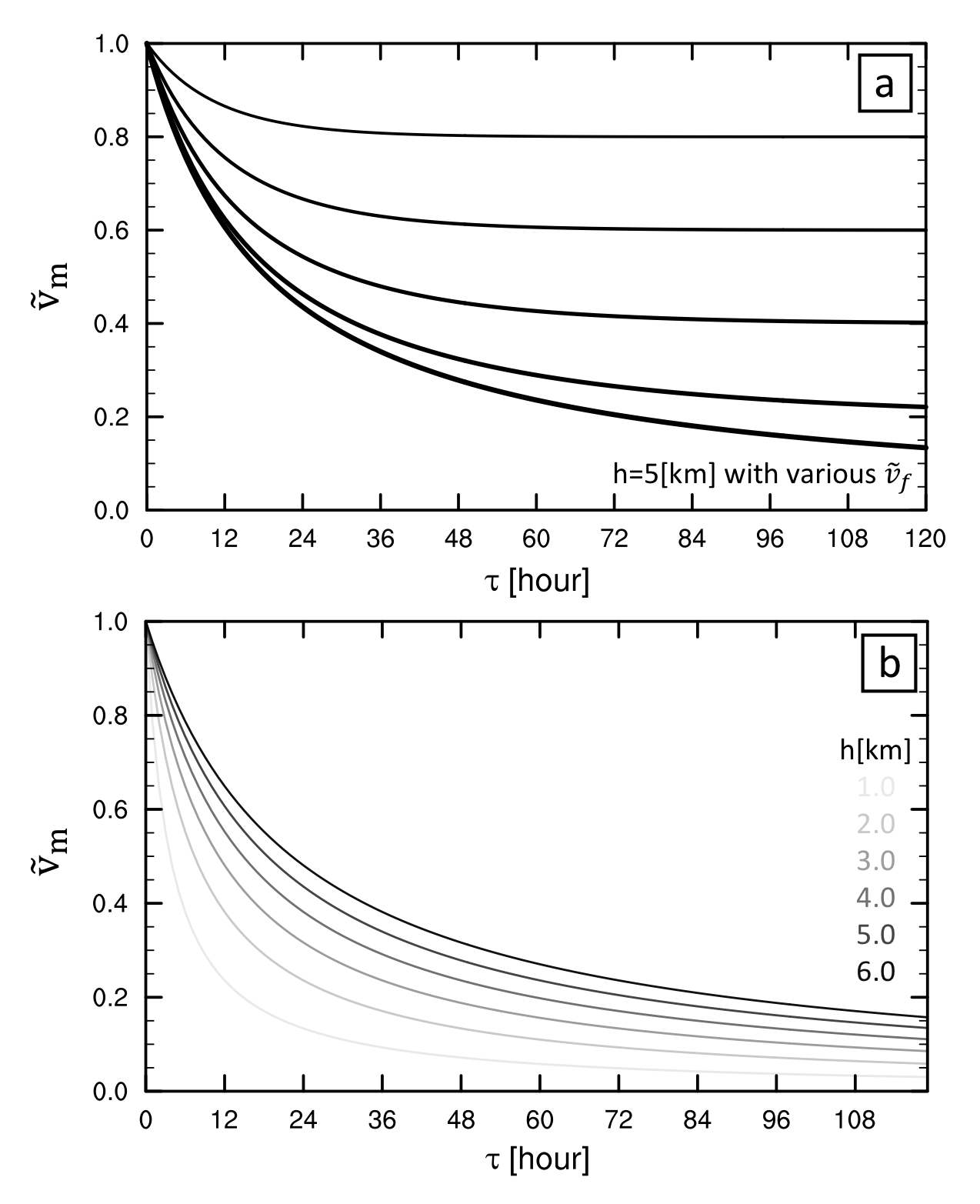}}
\caption{Normalized intensity decay predicted by E12 theory with $v_m(0)=100\;ms^{-1}$: (a) varying $\tilde v_f$ with $h=5km$, for $v_f=$80\%, 60\%, 40\%, 20\% and 0\% of initial intensity (Eq.\eqref{Eq_E12_decay1}); (b) varying $h$ with $\tilde v_f=0$ (Eq.\eqref{Eq_E12_vp0}). Note that for $\tilde v_f=0$, intensity reaches zero only as $\tau \rightarrow \infty$.}\label{E12_theoretical}
\end{figure}

E12 presents a theory for the time-dependent intensification of a TC, based on the role of outflow turbulence in setting the radial distribution of low-level entropy proposed in \cite{Emanuel+Rotunno2011}. The equation is given by  
\begin{equation}\label{Eq_E12}
v_{m}(\tau)=v_{f}\,\tanh{(\frac{C_k v_{f}}{2h}\tau)}
\end{equation}
with initial condition $v_m=0$ at $\tau=0$, where $v_{max}$ is the final steady-state maximum wind speed and $h$ is a constant boundary layer depth scale within the eyewall. \cite{Ramsay2020} generalized this equation to predict intensification from a non-zero initial intensity. Here we further generalize Eq.\eqref{Eq_E12} to represent the decay from an initial intensity $v_m(0)$ to a weaker equilibrium intensity $v_{f}$ (see Appendix). This can be written in a form normalized by the initial intensity given by
\begin{equation}\label{Eq_E12_decay1}
\tilde v_{m(th)}(\tau)=\tilde v_{f}\coth{\left(\frac{C_k v_{f}}{2h}\tau+\tanh^{-1}(\tilde v_f)\right)}
\end{equation}
where $\tilde v_{m(th)}(\tau)=\frac{v_{m(th)}(\tau)}{v_m(0)}$ and $\tilde v_f=\frac{v_f}{v_m(0)}$. We use the subscript $th$ to denote the intensity predicted by theory. Examples of the theoretical prediction of Eq.\eqref{Eq_E12_decay1} are presented in Fig.\ref{E12_theoretical}a for a range of $v_f$, using the value of $h$ (5 km) that was applied in E12 and an initial intensity of $100\; ms^{-1}$. 

When $v_{f}=0$, Eq.\eqref{Eq_E12_decay1} reduces to
\begin{equation}\label{Eq_E12_vp0}
\tilde v_{m(th)}(\tau)=\left(\frac{C_k v_m(0)}{2h}\tau +1\right)^{-1}
\end{equation}
For this special case, the transient intensity response from a given $v_m(0)$ depends only on the boundary layer depth scale $h$. Fig.\ref{E12_theoretical}b presents examples varying $h$ over a wide range of values, with using $v_m(0)=100\;ms^{-1}$ and $v_{f}=0$. Eq. \eqref{Eq_E12_vp0} is the continuous limit of Eq.\eqref{Eq_E12_decay1} as $v_{f}\rightarrow 0$ and is not singular. Note that setting $v_{f}=0$ produces a solution that does not actually reach zero intensity in finite time, but rather approaches zero only as $\tau \rightarrow \infty$; for realistic timescales, the solution still predicts a normalized intensity appreciably greater than zero ($\approx$15\% after 5 days in Fig.\ref{E12_theoretical}).

According to Eq.\eqref{Eq_E12_decay1}-\eqref{Eq_E12_vp0}, the intrinsic time scale of intensity decay from the initial condition is determined by the boundary layer depth scale $h$ and the steady-state final intensity $v_f$, taking $C_k$ as constant, such that a smaller $h$ leads to a faster decay (Fig.\ref{E12_theoretical}b). Effects from changes in the external environment, e.g., surface properties, are captured in the prediction of $v_{f}$ as discussed below. Notably, the precise definition of $h$ in E12 theory and its relation to the true boundary layer height, $H$, is uncertain both theoretically and practically. First and foremost, the TC boundary layer height $H$ is poorly understood even for storms over the ocean where the $H$ is approximated by its dynamical or thermodynamical characteristics \citep{Kepert2001,Emanuel1997,Bryan+Rotunno2009, Zhang2011, Seidel2010}. Unfortunately, these estimates of boundary layer heights can vary substantially from one another \citep{Zhang2011}. Moreover, in E12 theory it is the boundary layer depth specifically within the deeply-convecting eyewall that is relevant, where air rapidly rising out of the boundary layer effectively blurs the distinction between boundary layer and free troposphere \citep{Marks2008, Kepert2010, Smith+Montgomery2010}; perhaps for this reason E12 found a value corresponding to an approximate half-depth of the troposphere (5km) to perform best. Finally, little is known about the TC boundary layer height during the landfall transition. Thus, when evaluating E12 theory, we simply test a range of values for $h$ and examine the extent to which variations in the best-fit values of $h$ across experiments align with variations in estimates of the boundary layer.


\section{Methodology}\label{method}

Idealized numerical simulation experiments of landfall are used to test the theoretical predictions discussed above.

\subsection{Simulation setup}

The pronounced spatiotemporal heterogeneity in surface properties from storm to storm in real-world landfalls requires sophisticated land-surface and boundary layer parameterizations \citep{Cosby1984, Stull1988, Davis2008, Nolan2009, Jin2010}. However, experiments in axisymmetric geometry with a uniform environment and boundary forcing can reveal the fundamental responses of a mature TC to individual surface roughening or drying, as introduced in CC20. Thus, this work extends the individual forcing experiments of CC20 to simulations where the surface is simultaneously dried and roughened with varying magnitudes of each.

\begin{table}
\caption{Parameter values of the CTRL simulation.}
\label{geom}
\begin{tabular}{ |p{1.5cm} p{3.5cm} p{1.5cm}| }
\hline
Model& Name& Value \\
\hline
 $l_h$ &hor. mixing length & 750-m\\
 $l_{inf}$& asymptotic ver. mixing length&  100-m\\
$C_k$ & exchange coef. of enthalpy&0.0015\\
$C_d$ & exchange coef. of momentum&0.0015\\
$H_{domain}$&model height& 25-km \\
 $L_{domain}$&model radius&3000-km\\
\hline 
Environment& Name & Value\\
\hline 
$T_{ST}$& surface temperature &300-K\\
$T_{tpp}$ &tropopause temperature   &200-K \\
$Q_{cool}$& radiative cooling ($\theta$)& $1 K day^{-1}$\\
$f$&Coriolis Parameter& $5\times 10^{-5} s^{-1}$\\
\botline
\end{tabular}
\end{table}

All experiments are performed using the Bryan Cloud Model (CM1v19.8) \citep{Bryan+Fritsch2002} in axisymmetric geometry with same setup as \cite{Chen+Chavas2020}; model parameters are summarized in Table \ref{geom}. Dissipative heating is included. We first run a 200-day baseline experiment to allow a mature storm to reaches a statistical steady-state, from which we identify the most stable 15-day period. We then define the Control experiment (CTRL) as the ensemble-mean of five 10-day segments of the baseline experiment from this stable period whose start times are each one day apart. From each of the five CTRL ensemble member start times, we perform idealized landfall restart experiments by instantaneously modifying the surface wetness and/or roughness beneath the CTRL TC, which are then averaged into experimental ensembles analogous to the CTRL. Surface wetness is modified by decreasing the surface evaporative fraction $\epsilon$, which reduces the surface latent heat fluxes $F_{LH}$ through the decreased surface mixing ratio fluxes $F_{qv}$ in CM1 (sfcphys.F) as
\begin{equation}\label{Eq_qvflux}
F_{qv}= \epsilon s_{10}  C_q  \Delta q
\end{equation} 
\begin{equation}\label{Eq_LHflux}
F_{LH}= \rho L_v F_{qv}
\end{equation}
where $C_q$ is the exchange coefficients for the surface moisture; $s_{10}$ is the 10-m wind speed; and $\Delta q$ is the moisture disequilibrium between the 10-m layer and the sea surface. Surface roughness is modified by increasing the drag coefficient $C_d$, which modulates the surface roughness length $z_0$ and in turn the friction velocity $u^*$ for the surface log-layer in CM1 as
\begin{equation}\label{Eq_Cd_z0}
z_0= \frac{z}{e^{(\frac{\kappa}{\sqrt{C_d}}-1)}}
\end{equation}
\begin{equation}\label{Eq_Cd_ust}
u^*= max\left[\frac{\kappa s_1}{\ln(\frac{z_a}{z_0}+1)}, 1.0^{-6}\right]
\end{equation} where $\kappa$ is the von K\'arm\'an constant, $z=10m$ is the reference height, $s_1$ is the total wind speed on the lowest model, and $z_a$ is approximately equal to the lowest model level height. Readers are referred to CC20 for full details.

\begin{figure}[t]
\centerline{\includegraphics[width=0.5\textwidth]{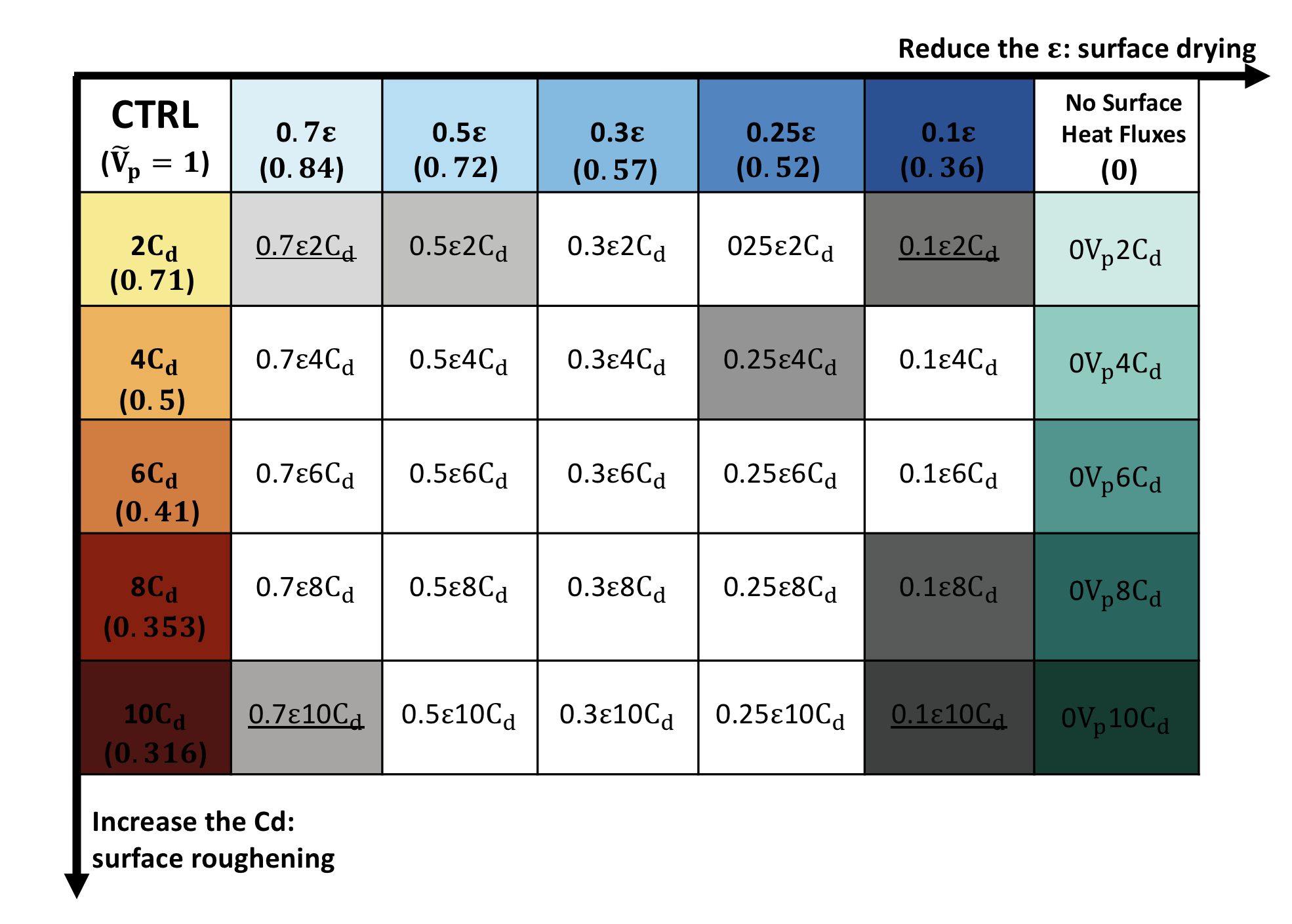}}
\caption{Two-dimensional experimental phase space of surface drying (decreasing $\epsilon$ moving left to right) and surface roughening (increasing $C_d$ moving top to bottom). CTRL is an ocean-like surface with $(C_d,\epsilon) = (0.0015,1)$. Values of the potential intensity response $\tilde V_p$ for CTRL and individual drying or roughening are listed in parentheses; $\tilde V_p$ for any combination of forcing is the product of $\tilde V_p$ for each individual forcing. Experiments testing combined forcings are shaded grey and the subset testing the most extreme combinations of each forcing are underlined. Experiment set $0V_pXC_d$, corresponding to the special case where surface heat fluxes are entirely removed ($V_p=0$), are shaded green.}\label{Table_experiments}
\end{figure}

Our experiments are summarized in Fig.\ref{Table_experiments}. Surface roughening-only experiments ($2C_d$, $4C_d$, $6C_d$, $8C_d$, $10C_d$) and drying-only experiments ($0.7\epsilon$, $0.5\epsilon$, $0.3\epsilon$, $0.25\epsilon$, $0.1\epsilon$) are fully introduced in CC20. The combined experiments are simulated in the same manner but with the surface dried and roughened simultaneously. Our combined experiments are designed in a way where individual drying and roughening are systematically paired with each other; experiments are named by the corresponding modifications in $C_d$ and $\epsilon$. We focus on two specific subsets of experiments within this phase space. First, $0.7\epsilon 2C_d$, $0.7\epsilon 10C_d$, $0.1\epsilon 2C_d$, $0.1\epsilon 10C_d$ are chosen as the representatives for extreme combinations, where each forcing takes its highest or lowest non-zero magnitude. Second, $0.25\epsilon 4C_d$, $0.5\epsilon 2C_d$, $0.1\epsilon 8C_d$ are chosen to represent cases where the individual forcings are varied in ways that yield comparable equilibrium potential intensities; $\tilde V_{p(\epsilon)}$ and $\tilde V_{p(C_d)}$ are labeled in Figure \ref{Table_experiments}. Finally, we generate a special set of combined experiments, $0V_pX C_d$, in which $\tilde V_p$ is fully reduced to zero for a range of magnitudes of roughening. Experiments are modeled by setting both surface sensible and latent heat fluxes to zero while increasing the roughness by a factor of $X$. This set of experiments sharing the same $\tilde V_p=0$ and are applied to test the simplified form of E12 assuming $\tilde v_f=0$ (Eq.\eqref{Eq_E12_vp0}).

\subsection{Testing theory against simulations}

In each experiment, the simulated storm intensity $v_{m}(\tau)$ is normalized by the time-dependent, quasi-stable CTRL value, where $\tau$ denotes the time since the start of a given forcing experiment. We primarily focus on the first 36-h evolution, during which $\tilde v_m$ decreases monotonically across all roughening, drying, and combined experiments. In addition, both the near-surface (50m) intensity response $\tilde v_{m(50m)}$ and above-BL (2km) intensity response $\tilde v_{m(2km)}$ are compared to theoretical predictions. The near-surface wind field is essential for predicting the inland TC hazards, while the above-BL wind field is generally used when formulating physically-based theories above the boundary layer. Over the ocean, $\tilde v_{m(50m)}$ and $\tilde v_{m(2km)}$ typically co-evolve closely \citep{Powell2003}. However, at and after landfall, the response of near-surface winds is expected to deviate from the above-BL winds, particularly in the case of roughening, as shown in CC20: there is a very rapid initial response of angular momentum to enhanced friction near the surface during the first 10 minutes (and especially the first 5-6 minutes) that subsequently propagates upward as the vortex decays. In contrast, the response to surface drying initially occurs aloft in the eyewall before propagating into the boundary layer, though this response is generally slower and smoother in time. Therefore, it is practically useful to test intensity theories against both near-surface and above-BL winds.

For E86 theory, we first compare the simulated equilibrium intensity against the equilibrium E86 prediction of Eq.\eqref{predictedVp}. We then compare the full time-dependent simulated intensity response against that predicted by assuming the total response is the product of the individual responses, $\tilde v^{*}(\tau)$ (Eq.\eqref{vm_deconstruct}). For E12 theory, we compare the simulated intensity evolution $\tilde v_m(\tau)$ against the E12 prediction $\tilde v_{m(th)}(\tau)$ of Eq.\eqref{Eq_E12_decay1}-Eq.\eqref{Eq_E12_vp0}. In these solutions, the initial intensity $v_m(0)=100.4\,ms^{-1}$ for 2km wind field and $v_m(0)=90.38\,ms^{-1}$ for 50m wind field, respectively. The final intensity $v_{f}$ is set as the minimum value of $v_{m}$ in each simulation during the evolution. A range of $h$ will be tested in Eq.\eqref{Eq_E12_decay1}, which will be discussed in the following subsection. For real-world landfalls, we do not know the minimum intensity prior to the inland evolution. Thus, as a final step, we compare simulation results against the theory with $\tilde v_f$ predicted from $\tilde V_p$. This final step may be useful for potentially applying the theory to real-world landfalls.

\subsection{The boundary layer depth scale $h$}

As introduced in Section 2, $h$ is an uncertain boundary layer height scale both theoretically, as it is assumed to be constant, and in practice, because we do not have a simple means of defining the boundary layer depth particularly during the landfall transition. Thus, we do not aim to resolve this uncertainty in $h$ in the context of landfall, but rather we simply test what values of $h$ provide the best predictions and evaluate to what extent variations in $h$ across experiments align with variations in estimates of the boundary layer height $H$. 

For each simulation, we test a range of constant values of $h$ from 1.0 to 6.0 km in 0.1 km increments in order to identify a best-fit boundary layer depth scale, $h_{BEST}$. We define $h_{BEST}$ as the value of $h$ that produces the smallest average error throughout the first 36-h evolution for each experiment. We then compare the systematic variation in $h_{BEST}$ against that of three typical estimates of boundary layer height $H$ calculated from each simulation. 

Since the E12 solution applies within the convecting eyewall region, we estimate $H$ using three typical approximations and measure the value at the radius of maximum wind speed ($r_{max}$) at the lowest model level: 1. $H_{v_m}$ is the height of maximum tangential wind speed \citep{Bryan+Rotunno2009}; 2. $H_{inflow}$ is the height where radial wind $u$ in the eyewall first decreases to 10\% of its surface value \citep{Zhang2011}; and 3. $H_{\theta_v}$ is defined as the height where $\theta_v$ in the eyewall matches its value at the lowest model level \citep{Seidel2010}. The 36-h evolution of each estimate of $H$ across our simulations is shown in Supplementary Figure 1. Considering that $H$ varies in time during the experiment, the averaged value of $H$ during $\tau=0-6\;h$ and $\tau=30-36\;h$ are each compared to the $h_{BEST}$.

\section{Results}\label{results}

\subsection{Near-surface vs. above-BL intensity response}
As discussed in Section 3b, the simulated intensity responses near the surface ($z=50m$) and above the boundary layer ($z=2km$) may differ, and thus we intend to test the theory against both. We begin by simply identifying important differences between the intensity responses at each level. 

\begin{figure*}[t]
\centerline{\includegraphics[width=0.9\textwidth]{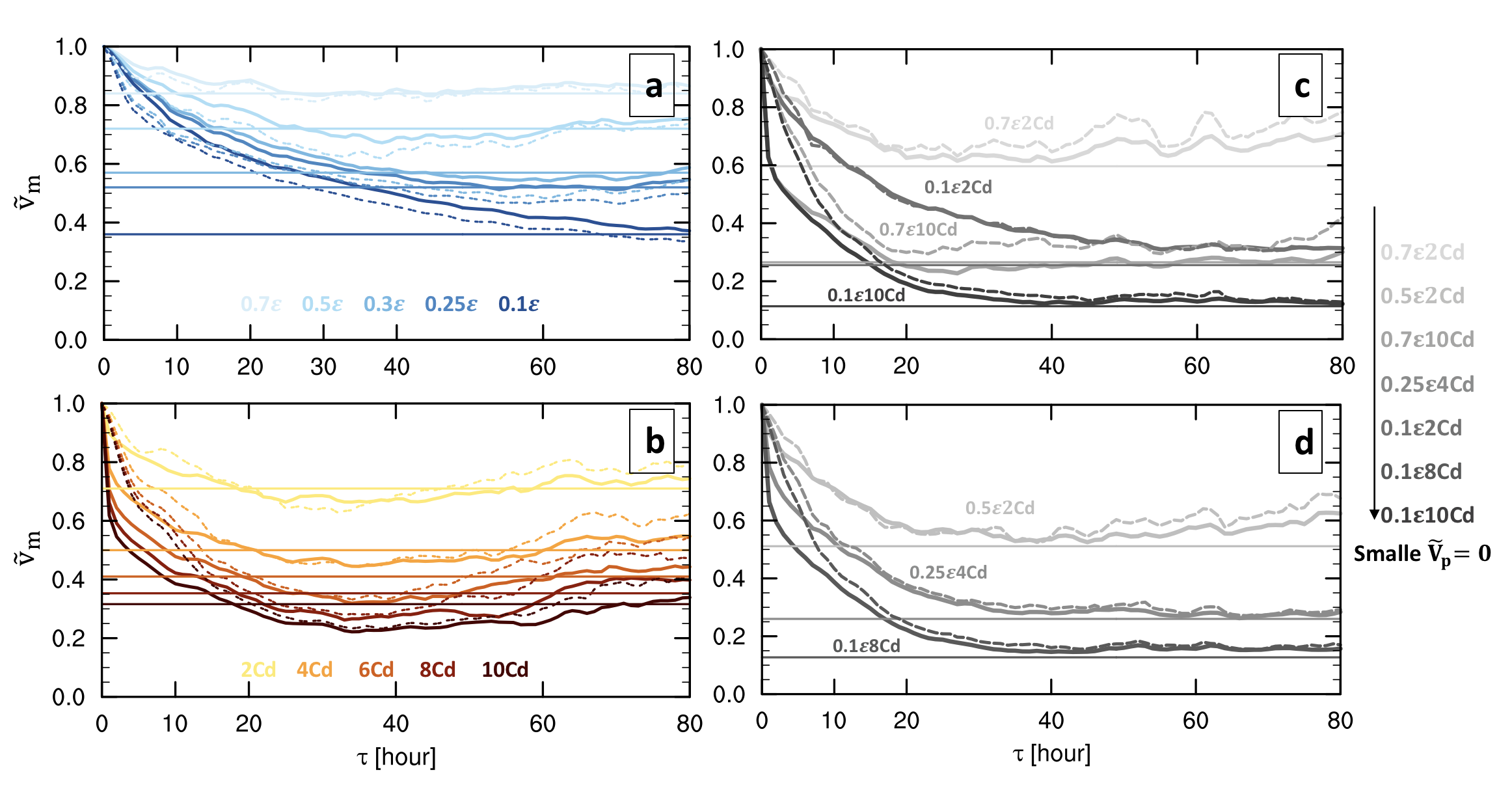}}
\caption{Temporal evolution of simulated near-surface (50m, solid curves) and above-BL (2km, dash curves) $\tilde v_m$, across experiments: (a) surface drying, (b) surface roughening and (c)-(d) representative combined experiments. Potential intensity response, $\tilde V_p$ (Eq.\eqref{Vp_separate2}-\eqref{predictedVp}), denoted by horizontal line. Darker color indicates a smaller $\tilde V_p$.}\label{intensity-All}
\end{figure*}

For drying-only experiments, $\tilde v_{m(2km)}$ responds to reduced $\epsilon$ before $\tilde v_{m(50m)}$ during the initial $\sim 5$ hours (Fig.\ref{intensity-All}a), where stronger drying results in a larger deviation between the responses at each level. $\tilde v_{m(2km)}$ decreases slightly more rapidly than $\tilde v_{m(50m)}$ but they eventually converge to comparable equilibria. In contrast, for roughening $\tilde v_{m(50m)}$ responds nearly instantaneously across all roughening experiments, whereas $\tilde v_{m(2km)}$ responds more gradually. The magnitude of the deviation of $\tilde v_{m(50m)}$ from $\tilde v_{m(2km)}$ during the first 20 hours increases with increasing roughening (Fig.\ref{intensity-All}b), though again both converge to comparable values after $\tau=20\;h$. The above behavior is consistent with the findings in CC20, where surface roughening has a significant and immediate impact on near-surface intensity while drying first impacts the eyewall aloft. As noted in CC20, $\tilde V_p$ provides a reasonable prediction for the long-term equilibrium intensity response to each individual forcing; there is a small overshoot in roughening experiments where $\tilde v_m$ reaches a minimum that is less than $\tilde V_p$ at approximately $\tau=36\; h$ before increasingly gradually towards $\tilde V_p$. 

For our combined experiment sets (Fig.\ref{intensity-All}c-d), both surface drying and roughening determine the total response of storm intensity. However, regardless of the relative strength of each forcing, $\tilde v_{m(50m)}$ always decreases more rapidly than $\tilde v_{m(2km)}$ due to the surface roughening. Similar to the roughening-only experiment, stronger roughening results in a larger deviation of $\tilde v_{m(50m)}$ from $\tilde v_{m(2km)}$ during the first 20 hours. Overall, the rapid initial response of near-surface intensity is controlled by the surface roughening regardless of the surface drying magnitudes. Moreover, similar to the individual forcing experiments, $\tilde V_p$ provides a reasonable prediction for the simulated minimum $\tilde v_m$ in combined experiments used to define the theoretical final intensity $\tilde v_f$.  

\subsection{Deconstructing simultaneous drying and roughening}\label{sec_deconstruct}
\begin{figure*}[t]
\centerline{\includegraphics[width=0.7\textwidth]{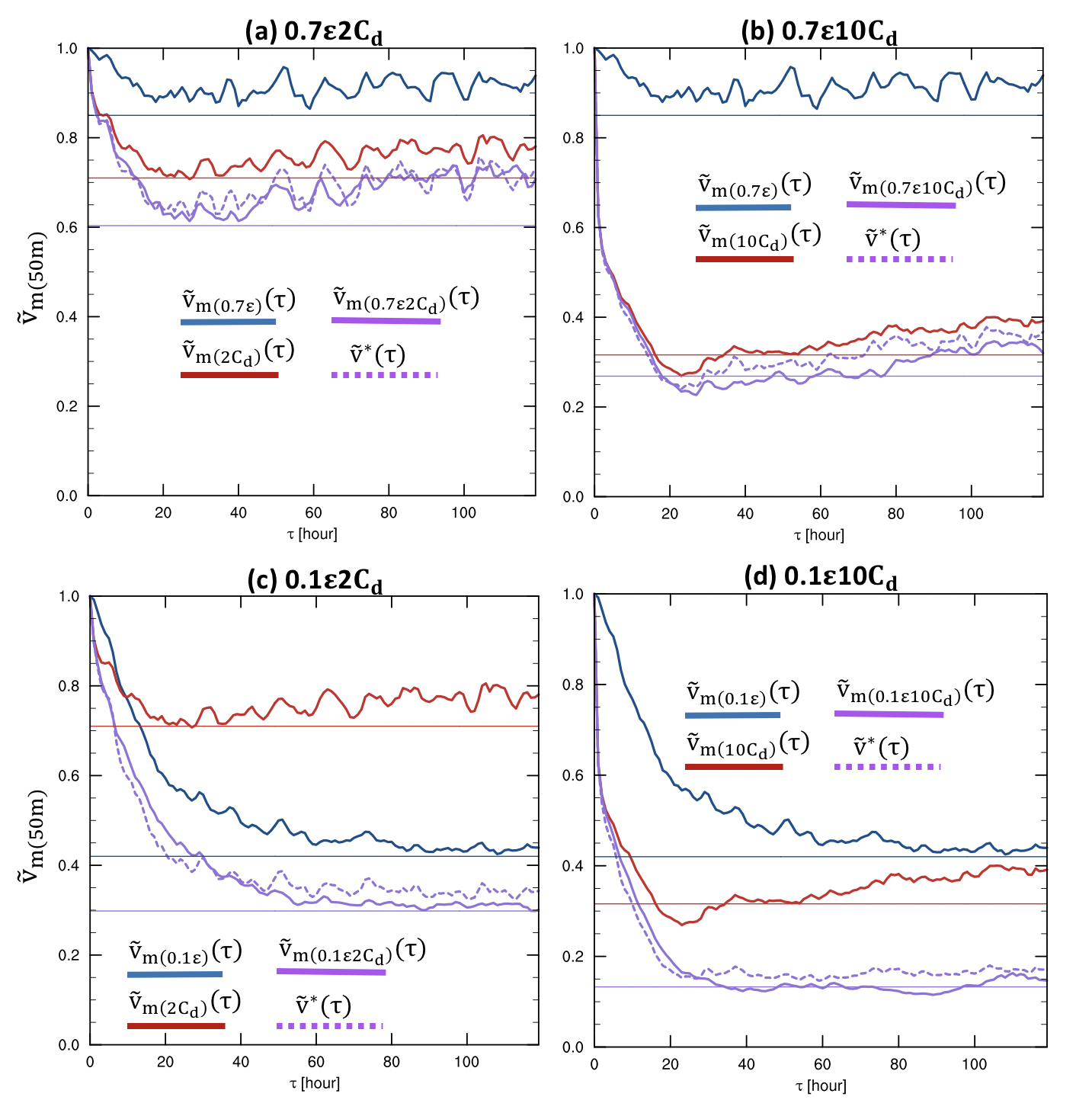}}
\caption{Temporal evolution of simulated $\tilde v_m(\tau)$ for four representative combined experiments (solid purple) and their associated individual forcing experiments (solid blue for drying, solid red for roughening), with prediction $\tilde v^{*}(\tau)$ (dash purple; Eq.\eqref{vm_deconstruct}) defined as the product of the individual forcing responses. Horizontal line denotes $\tilde V_p$, colored by experiment.}\label{E86combine}
\end{figure*}

Now we focus on the full time-dependent responses of the combined forcing experiments. As noted above, we hypothesized based on traditional potential intensity theory (Eq.\eqref{Vp_separate1}-\eqref{Vp_deconstruct}) that the transient response of storm intensity to simultaneous drying and roughening can be predicted as the product of their individual responses (Eq.\eqref{vm_deconstruct}). Thus, we compare the $\tilde v^{*}(\tau)$ to the simulated $\tilde v_m(\tau)$ of near-surface winds for each combined experiment (Fig.\ref{E86combine}) for our experiment set with extreme cases of combined drying and roughening. Regardless of the magnitude of roughening and/or drying, $\tilde v^{*}(\tau)$ follows the simulated $\tilde v_m(\tau)$ closely through the initial rapid decay forced by roughening, the weakening stage through 36 hours, and the final equilibrium stage. There is a slight low bias in $\tilde v^{*}(\tau)$ relative to the $\tilde v_m(\tau)$ throughout the primary weakening stage, especially for strong drying (Fig.\ref{E86combine}c-d), indicating that there is a slight compensation in the response to roughening when strong drying is also applied. Overall, though, the full temporal evolution of the normalized responses to drying and roughening can indeed be combined multiplicatively. A very similar result is obtained for the above-BL intensity as well (not shown).  

Although E86 theory is formulated for the equilibrium intensity, the above results indicate that the implication of its underlying physics also extends to the transient response to simultaneous surface drying and roughening. This behavior aligns with the notion that periods of intensity change represent a non-linear transition of the TC system between two equilibrium stable attractors given by the pre-forcing and post-forcing $V_p$ \citep{Kieu+Moon2016, Kieu+Wang2017}. Because the distance between attractors is multiplicative, evidently so too is the trajectory between them.

\subsection{Testing theory for transient intensity response}

\begin{figure*}[t]
\centerline{\includegraphics[width=0.8\textwidth]{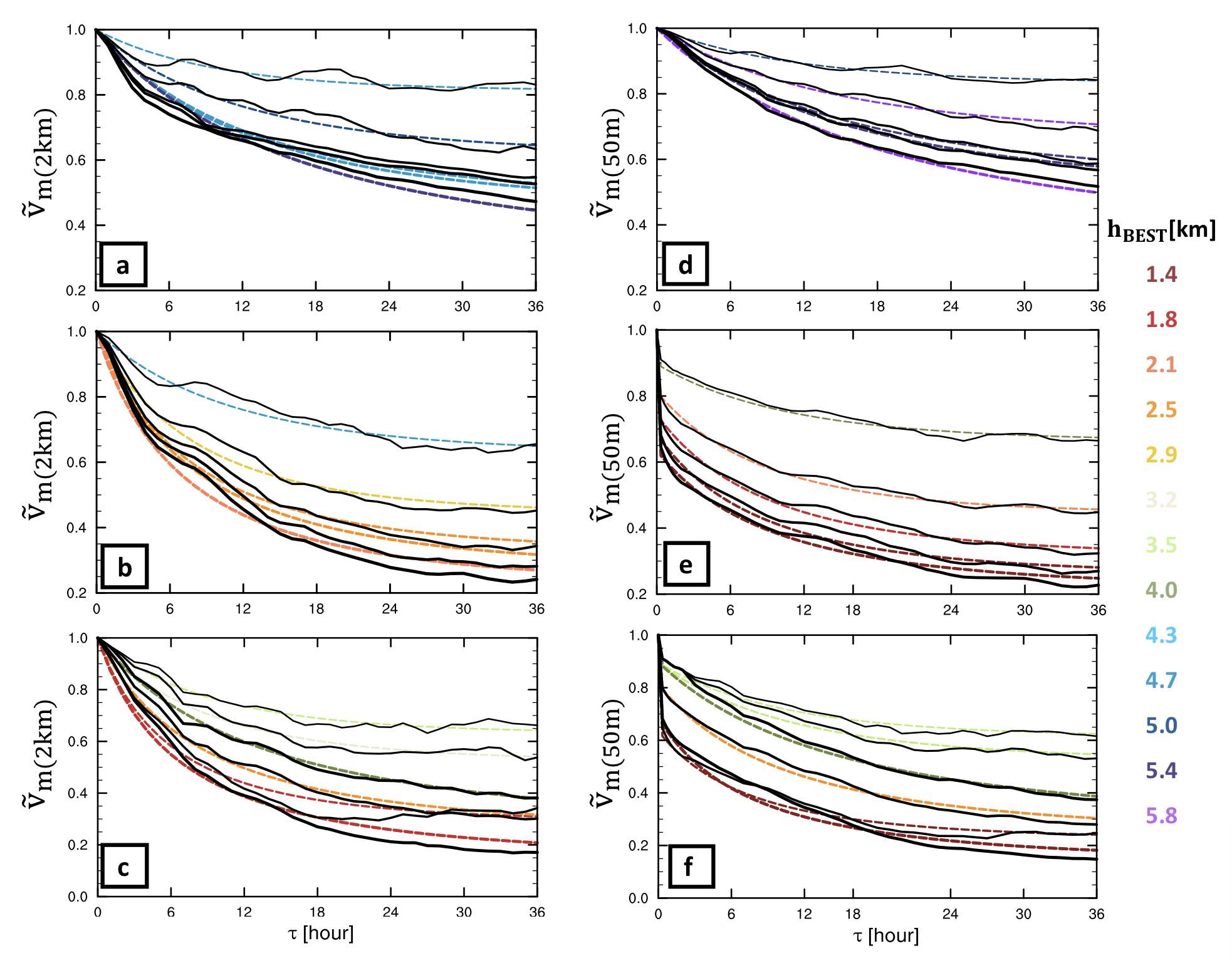}}
\caption{ (a-c) Temporal evolution of the simulated above-BL (2km) winds $\tilde v_{m}(\tau)$ (solid black) and theoretical prediction $\tilde v_{m(th)}(\tau)$ (dash colored; Eq.\eqref{Eq_E12_decay1}) for (a) drying, (b) roughening and (c) combined experiments. $\tilde v_{m(th)}(\tau)$ is colored by the value of best-fit value of $h$, $h_{BEST}$. (d-f) are same as (a-c) but for the near-surface (50m) wind, in which (e-f) are the predictions from Eq.\eqref{Eq_E12_decay2} to account for the initial rapid decay due to roughening.}\label{Best-Hall1}
\end{figure*}

We next test the extent to which E12 theory (Eq.\eqref{Eq_E12_decay1} and Eq.\eqref{Eq_E12_vp0}) can predict the intensity evolution, taking as input $\tilde v_{f}$ and a best-fit value of $h_{BEST}$ from each simulation. The subsequent subsection compares the systematic variation in $h_{BEST}$ against that of the estimated $H$ in each set of experiments. 

We begin by focusing on the above-BL intensity. The E12-based solution (Eq. \eqref{Eq_E12_decay1}) can reasonably capture the overall transient intensity response across all roughening, drying, and combined experiments (Fig.\ref{Best-Hall1}a-c). For drying-only experiments (Fig.\ref{Best-Hall1}a), the theory initially ($\tau<5\;h$) underestimates $\tilde v_{m(2km)}$ (Fig.\ref{intensity-All}a). $h_{BEST}$ in drying experiments takes a value around 5 km and shows little systematic variations with increased drying magnitude (Fig.\ref{Best-Hall1}a). In contrast, $h_{BEST}$ decreases with increased surface roughening, from 4.3 km to 2.1 km (Fig.\ref{Best-Hall1}b). In the following subsection, we compare this variation in $h_{BEST}$ to that of multiple estimates of $H$.

For near-surface winds, E12 theory can capture the transient response of $\tilde v_m$ in drying experiments (Fig.\ref{Best-Hall1}d) using a slightly larger $h_{BEST}$. For roughening and combined experiments, though, it misses the very rapid initial decay due to enhanced $C_d$ (Fig.\ref{intensity-All}b-d and Fig.\ref{Best-Hall1}e-f), which was shown in Fig.1 of CC20 to occur within the first 10 minutes. The magnitude of this rapid initial decay depends on $C_d$, as surface roughening immediately acts to rapidly remove momentum near the surface first. To account for this initial response, we propose a simple model for this initial decay of $\tilde v_{m(50m)}$ that may be derived from the tangential momentum equation in a 1D slab boundary layer with a depth of $h^{*}$, given by 
\begin{equation}\label{Eq_KEbudget1}
h^{*}\frac{dv}{dt}=-C_d \left| \vb{v} \right|v
\end{equation}
Taking $\left| \vb{v}\right| \approx v$, this equation may be integrated and then normalized by an initial intensity $v_m(0)$ to yield 
\begin{equation}\label{Eq_KEbudget2}
\tilde v_{m(th)}(\tau)=\left(\frac{C_d v_m(0)}{h^{*}}\tau+1\right)^{-1}
\end{equation}
Curiously, Eq.\eqref{Eq_KEbudget2} is mathematically identical to the E12 solution with $v_f = 0$ (Eq.\eqref{Eq_E12_vp0}) except with $C_k$ replaced by $2C_d$, a topic we return to in the summary section.

We may combine Eq.\eqref{Eq_KEbudget2} for the first 10-min evolution with Eq.\eqref{Eq_E12_decay1} to model the complete near-surface intensity response to surface roughening: 
\begin{equation}\label{Eq_E12_decay2}
 \tilde v_{m(th)}(\tau)=\bigg\{
\begin{array}{c c}	
     \left(\frac{C_d v_m(0)}{h^{*}}\tau +1\right)^{-1}& \tau \leq 10\;min\\
      \tilde v_{f}\,\coth{(\frac{C_k v_{f}}{2h}\tau+\tanh^{-1}(\tilde v_{f}^{*}))}& \tau> 10\;min
\end{array}
\end{equation}
where $\tilde v_{f}^{*}=\frac{v_f}{v_{m(th)}(10\;min)}$ since $v_{m(th)}$ now decreases from a new initial intensity $v_{m(th)}(10\;min)$ calculated from the first equation. For the $\tau\le10 \; min$ solution, we find that the rapid decay by $\tau=10\;min$ across all roughening experiments can be captured by setting $h^{*} = 1.78\;km$ constant, which yields a constant decay rate during the period $\tau=0-10$ min. In reality, the decay rate is very large in the first minute and monotonically decreases through the 10 minute period. This time-varying decay rate can be reproduced by allowing $h^{*}$ to increase with time from an initial very small value, which aligns with the physical response to roughening that may be thought of as the formation of a new internal boundary layer that begins at the surface and rapidly expands upward. Here though we employ a constant $h^*$ in order to retain a simple analytic solution; the 10-minute period is short enough that the difference is likely not of practical significance. Thereafter, we estimate $h_{BEST}$ for each simulation in the same manner as before.

The comparison of model against simulations is shown in Fig.\ref{Best-Hall1}e-f.  Eq.\eqref{Eq_E12_decay2} can capture the simulated near-surface transient intensity response $\tilde v_m(\tau)$ across roughening and combined experiments. The values of $h_{BEST}$ for the prediction of $\tilde v_{m(50m)}$ are similar to those obtained for the above-BL case but slightly smaller in magnitude (Fig.\ref{Best-Hall1}b,e and c,f). 

\begin{figure}[t]
\centerline{\includegraphics[width=0.45\textwidth]{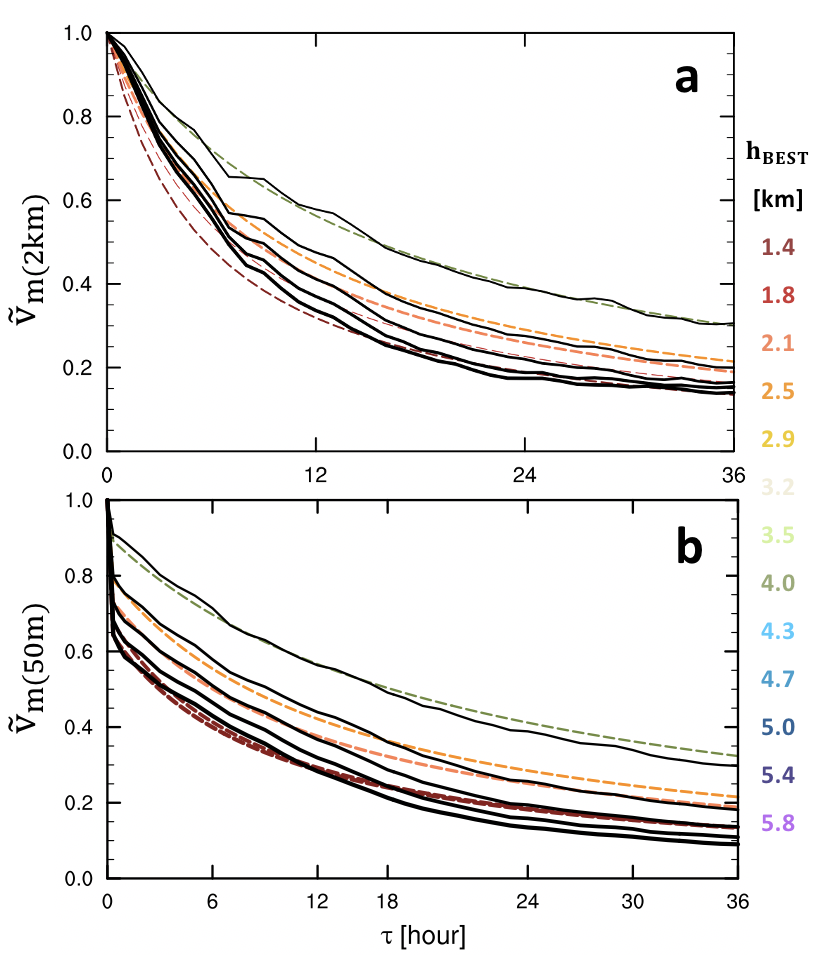}}
\caption{Temporal evolution of simulated intensity response and theoretical prediction for the experiment set $0V_pXC_d$: (a) above-BL 2km winds (Eq.\eqref{Eq_E12_vp0}) and (b) near-surface 50m winds (Eq.\eqref{Eq_E12_Vp0-2}). E12-based prediction is colored by the value of $h_{BEST}$ for corresponding experiment.} \label{Vp0_E12}
\end{figure}

Similar behavior associated with roughening is found in experiments for the special case $v_f=0$ (Fig.\ref{Vp0_E12}). Thus, we again propose a two-stage model for $\tilde v_{m(50m)}$ given by
\begin{equation}\label{Eq_E12_Vp0-2}
\tilde v_{m(th)}(\tau)=\bigg\{
\begin{array}{c c}	
    \left(\frac{C_d v_m(0)}{h^{*}}\tau +1\right)^{-1}& \tau \leq 10\;min\\
     \left(\frac{C_k v_m(0)}{2h}\tau +\frac{v_m(0)}{v_{m(th)}(10\;min)}\right)^{-1}& \tau >10\;min
\end{array}
\end{equation}
The comparison of model against simulations is shown in Fig.\ref{Vp0_E12}, where $h_{BEST}$ exhibits a decreasing trend with enhanced surface roughening, similar to that found in the pure roughening experiments. 

Note that multiplying both sides of \eqref{Eq_KEbudget2} by $v$ yields a budget equation for kinetic energy given by $h\frac{d(KE)}{dt}=-C_d v^3$, where $KE=\frac{1}{2}v^2$ and the RHS is the expression for surface frictional dissipation of kinetic energy that is standard in TC theory \citep{Bister1998,Tang+Emanuel2010, Chavas2017}\footnote{Note: the KE budget equation may be expressed as $h\frac{d(KE)}{dt}=-2C_d |\textbf{v}|(KE)$. Hence, $2C_d$ represents the surface exchange coefficient of kinetic energy over a static surface.}. Physically, then, the solution for the initial roughening response represents the intensity response to the dominant sink of kinetic energy in the absence of the dominant compensating thermodynamic source of kinetic energy from surface heat fluxes for a tropical cyclone. After this initial response, the solution follows a solution that accounts for both source and sink as encoded in E12 theory. The interpretation is that there exists a brief initial period where surface roughening directly modifies the near-surface air in a manner that is thermodynamically independent of the rest of the vortex. Thereafter, the vortex has adjusted and weakening proceeds according to processes governed by the full TC system, analogous to that of drying. In reality this transition is likely not instantaneous, though we have modeled it as so here for simplicity. The details of this adjustment process warrant more in-depth investigation that is left for future work.

\subsubsection{Comparing the variation of $h_{BEST}$ and estimates of $H$}
\begin{figure*}[t]
\centerline{\includegraphics[width=0.75\textwidth]{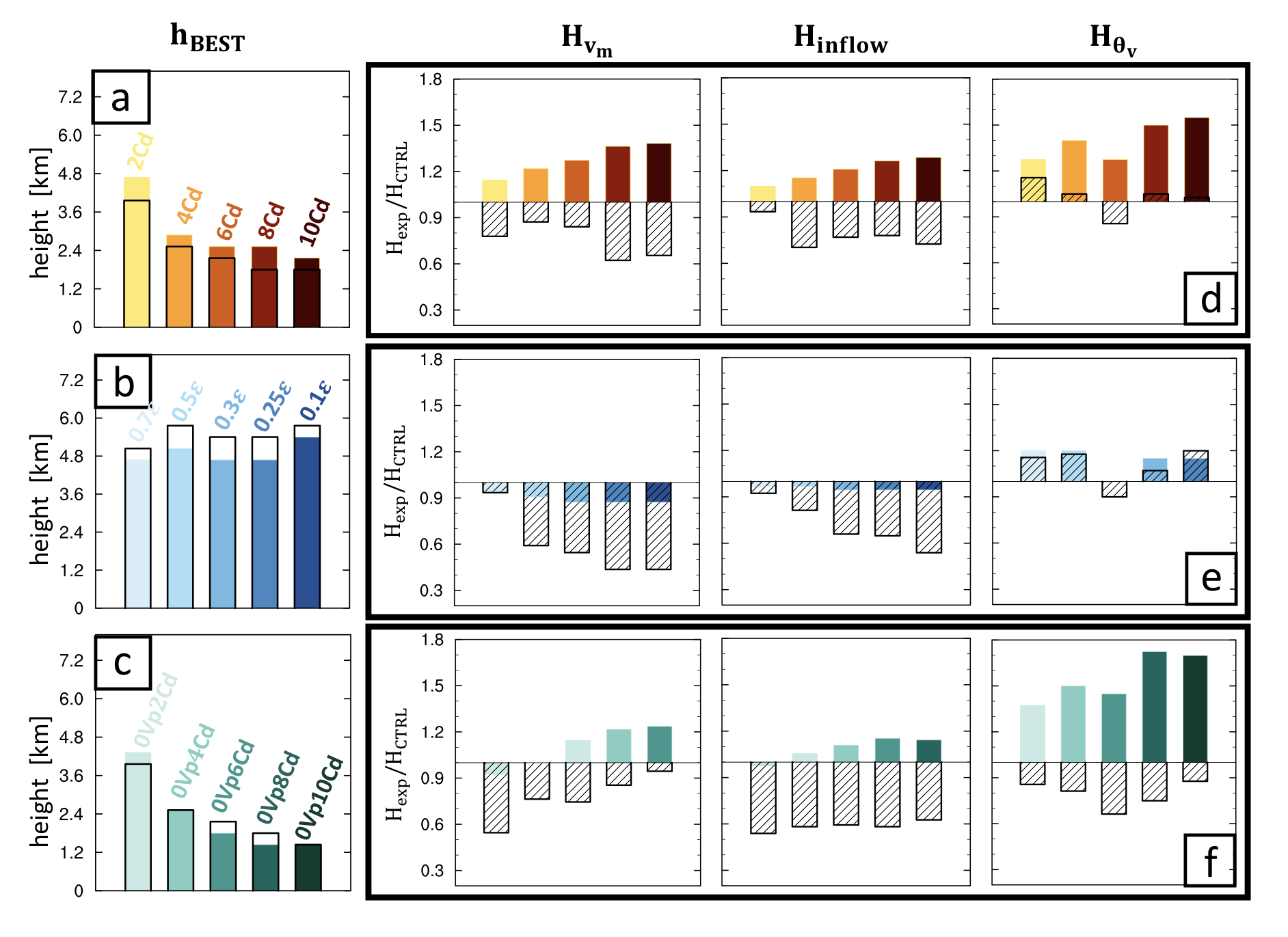}}
\caption{Comparison of $h_{BEST}$ (a-c) and the systematic variation of three estimates of boundary layer depth $H$ (d-f) for each set of surface roughening, surface drying, and $0V_pXC_d$ experiments. For (a-c), $h_{BEST}$ is shown for above-BL (color) and near-surface (box) intensity predictions. For (d-f), $\frac{H_{exp}}{H_{CTRL}}$ is shown for the early period ($\tau=0-6h$; color) and later period ($\tau=30-36h$; hatched box). The CTRL values of $H_{v_m}$, $H_{inflow}$, and $H_{\theta_v}$ are 0.95km, 1.56km, and 0.67km, respectively.} \label{compareH}
\end{figure*}

We next compare the variations of $h_{BEST}$ from our simulations to that of three common estimates of the boundary layer height: $H_{v_m}$, $H_{inflow}$, and $H_{\theta_v}$. We compare trends for both the initial response (0-6h) and the equilibrium response (30-36h) (Fig.\ref{compareH}d-f). As noted above, $h_{BEST}$ was found to decrease with increased roughening (Fig.\ref{compareH}a) but remain relatively constant for increased drying (Fig.\ref{compareH}b). Moreover, $h_{BEST}$ values are quite similar when estimated from the above-BL vs. near-surface responses (Fig.\ref{compareH}a-c), and hence this discussion is not dependent on the choice of level for defining $h_{BEST}$.

To show the variation of estimates of $H$ in different sets of experiments, we normalize each $H$ by its corresponding value in the CTRL experiment as $\frac{H_{exp}}{H_{CTRL}}$ (Fig.\ref{compareH}d-f), where each estimate of $H$ in the CTRL experiment is quasi-stable (Supplementary Fig.1). We focus first on the responses to pure roughening or pure drying (Fig.\ref{compareH}, top two rows). During 0-6h, all three estimates of $H$ slightly increase with enhanced roughening and are approximately constant with enhanced drying (Fig.\ref{compareH}d-f, color). By 30-36h, $H$ decreases with enhanced roughening and enhanced drying for $H_{v_m}$ and $H_{inflow}$, while $H_{\theta_v}$ remains relatively constant for each (Fig.\ref{compareH}d-f, shaded).


Overall, there is no single estimated $H$ whose systematic variation matches that of $h_{BEST}$ (Fig.\ref{compareH}). For roughening-only experiments, the initial response of all three estimates of $H$ with enhanced roughening is opposite to that of $h_{BEST}$ (Fig.\ref{compareH}a,d), though the slow response of each $H$ for two of the estimates does show a decrease with enhanced roughening (Fig.\ref{compareH}d, shaded). For drying-only experiments, the initial response of $H$ is either constant or very slowly decreasing with enhanced drying (Fig.\ref{compareH}b,e), similar to $h_{BEST}$. However, the slow response of $H_{v_m}$ and $H_{inflow}$ decreases with enhanced drying in contrast to the nearly constant $h_{BEST}$, while the slow response of $H_{\theta_v}$ is again relatively constant (Fig.\ref{compareH}e, shaded).

Finally, for $0V_pXC_d$ experiments (Fig.\ref{Vp0_E12}), $h_{BEST}$ exhibits a similar systematic response to enhanced surface roughening as the pure roughening experiments (Fig.\ref{compareH} a,c). Turning off all heat fluxes does not significantly alter the variation of $h_{BEST}$ with roughening relative to the pure-roughening experiments; this behavior is also similar to the pure drying experiments where $h_{BEST}$ remains constant with enhanced drying. The systematic variation in $h_{BEST}$ disagrees with both the early and the slow response of all three estimates of $H$ (Fig.\ref{compareH} c,f).

Note that for the combined experiments, there is no clear evidence to link the decreasing trend in $h_{BEST}$ (Fig.\ref{Best-Hall1}c,f) to the change in each individual forcing. Thus, we elect not to speculate on the details of $h_{BEST}$ and $H$ for the combined experiments; instead, we explore a more practical theoretical prediction for combined forcing cases that applies $h_{BEST}$ to drying and roughening experiments individually as shown in the next section.


The disagreement in the systematic trends both among estimates of $H$ and between those estimates $H$ and $h_{BEST}$ motivates the need for more detailed studies on the TC boundary layer during and after the landfall in future work. In terms of E12 theory, though the solution can reproduce the decay evolution, it also likely oversimplifies the TC boundary layer, where a constant $h$ cannot fully capture the time- and the radially-varying response of boundary layer height to landfall-like surface forcing. In terms of boundary layer theory, the optimal definition of boundary layer depth is itself uncertain. Meanwhile, without a comprehensive understanding of the TC boundary layer during the landfall, it is unclear if one particular definition of $H$, if any, might be most appropriate for E12 theory within the eyewall. Therefore, having a better estimation of $h$ in the E12 solution and an improved understanding of boundary layer evolution during the landfall would help explain the differences in systematic variations between $h$ and estimates of $H$. 

\subsection{Predicting the intensity response combining equilibrium and transient theory}
We may combine all theoretical findings presented above to predict the near-surface intensity response to idealized landfalls for experiments combining drying and roughening. This represents the most complex of our experimental outcomes. We begin from the result of Eq.\eqref{vm_deconstruct} and predict the intensity response to simultaneous surface drying and roughening as the product of their individual predicted intensity responses:
\begin{equation}\label{E12_and_E86combine}
\tilde v_{th(C_d\epsilon)}\approx \tilde v_{th}^{*}=\tilde v_{th(C_d)} \tilde v_{th(\epsilon)}
\end{equation}
where $\tilde v_{th(\epsilon)}$ is generated by Eq.\eqref{Eq_E12_decay1} and $\tilde v_{th(C_d)}$ is generated by Eq.\eqref{Eq_E12_decay2}. When applying the transient solution for each individual forcing component, we use the previously-identified $h_{BEST}$ from Fig.\ref{compareH}a-b and predict the $\tilde v_f$ using $\tilde V_p$ given by Eq.\eqref{Vp_separate2}-\eqref{Vp_separate1}. In principle, an empirical model for $h_{BEST}$ for each individual forcing could be generated from our data since $h_{BEST}$ is approximately constant with enhanced drying and monotonically decreasing with enhanced roughening. But given the uncertainties in $h_{BEST}$ as a true physical parameter, we elect not to take such a step here.

\begin{figure}[t]
\centerline{\includegraphics[width=0.5\textwidth]{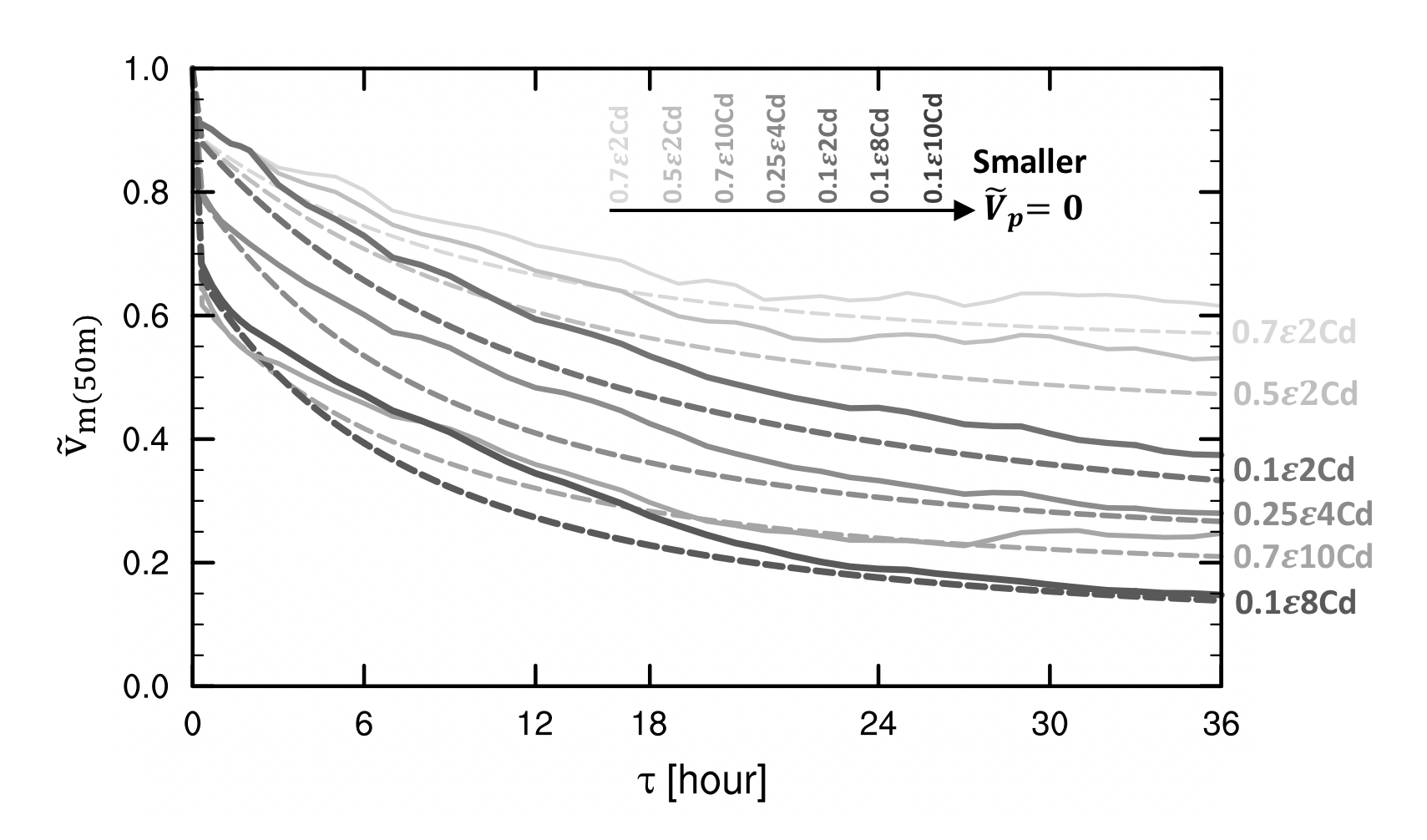}}
\caption{Temporal evolution of simulated near-surface intensity response for combined forcing experiments (solid) and the corresponding prediction combining the equilibrium and transient theory (dash; Eq.\eqref{E12_and_E86combine}), with $\tilde v_f = \tilde V_p$ and setting $h$ equal to the values of $h_{BEST}$ for the individual predicted responses to drying (Eq.\eqref{Eq_E12_decay1}) and roughening (Eq.\eqref{Eq_E12_decay2}).} \label{E12incorpE86}
\end{figure}

The results are shown in Fig.\ref{E12incorpE86}. Overall, our analytic theory performs well in capturing the first-order response across experiments, particularly given the relative simplicity of the method. There is a slight low bias in $\tilde v_{th(C_d\epsilon)}$ (i.e., too strong of a response) across all the predictions relative to the simulations, a reflection of the slight compensation found in the assumption that the combined response may be modeled as the product of the individual responses in Fig.\ref{E86combine}. Therefore, knowing how each surface forcing is changed after landfall provides a theoretical time-dependent intensity response prediction when given estimates of $\tilde V_p$ and $h_{BEST}$ for any combination of uniform surface forcing. This offers an avenue to link our theoretical understanding to real-world landfalls. 

\subsection{Comparison with existing empirical decay model}
Finally, as a first step toward linking theory and real-world landfalls, we provide a simple comparison of our theoretical model with the prevailing empirical model for inland decay. The model was first introduced by \cite{DeMaria+Kaplan1995} and most recently applied to historical observations by \cite{Jing+lin2019} as 
\begin{equation}\label{Eq_empirical}
V(t)=V_b+(V_0-V_b)e^{-\alpha t}
\end{equation}
where storm landfall intensity $V_0$ decays to a background intensity $V_b$ with a constant exponential decay rate $\alpha$. \cite{Jing+lin2019} estimated $V_b=18.82\;kt \; (9.7\;ms^{-1})$ and $\alpha=0.049\;h^{-1}$ using the historical Atlantic hurricane database.

\begin{figure*}[t]
\centerline{\includegraphics[width=0.75\textwidth]{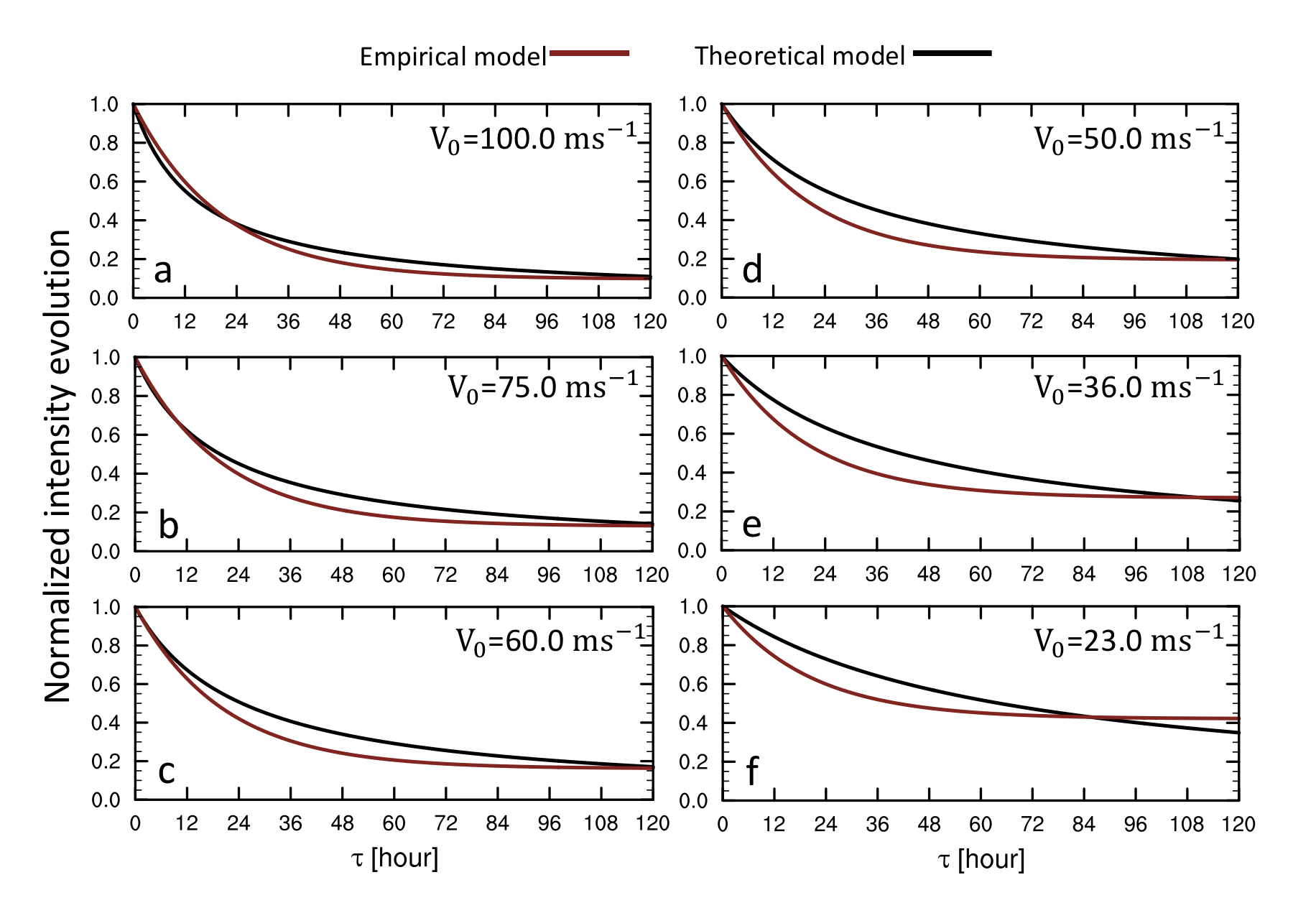}}
\caption{Comparison between the theoretical intensity prediction for $v_f=0$ (Eq.\eqref{Eq_E12_vp0}) against the prevailing empirical exponential decay model prediction from \cite{Jing+lin2019} (Eq.\eqref{Eq_empirical}) for a range of initial intensities ([100, 75, 60, 50, 36, 23] $ms^{-1}$). For the theoretical prediction, $h$=5km.} 
\label{emp_model}
\end{figure*}

Comparisons between theory and Eq.\eqref{Eq_empirical} predictions are shown in Fig.\ref{emp_model}, for $V_0$ from $100\;ms^{-1}$ to $23\;ms^{-1}$ similar to \cite{Jing+lin2019} (their Fig.3). Given that TCs in nature eventually dissipate (and typically do so rapidly), we choose the theoretical equation with $v_f=0$ (Eq.\eqref{Eq_E12_vp0}); this is also by far the simplest choice as no final intensity information is required. We set $h=5km$ constant. Eq.\eqref{Eq_E12_vp0} compares well against the empirical prediction for inland intensity decay, capturing the first-order structure of the characteristic response found in real-world storms. The comparison is poorer for weaker cases (Fig.\ref{emp_model}e-f), though uncertainties in intensity estimates are also likely highest for such weak storms. Note that the structure of the empirical model is constrained strongly by the assumption of an exponential model, so differences beyond the gross structure should not be overinterpreted. Ultimately, the consistency with the empirical model provides additional evidence that the physical model may indeed be applicable to the real world. Hence, it may help us to develop a physically-based understanding of the evolution of the TC after landfall.


\section{Summary}\label{summary}

This work tests the extent to which equilibrium and transient tropical cyclone intensity theory, the latter reformulated here to apply to inland intensity decay, can predict the simulated equilibrium and transient intensity response of a mature tropical cyclone to surface drying, roughening, and their combination. This work builds off of the mechanistic study of Chen and Chavas (2020) that analyzed the responses of a mature tropical cyclone to these surface forcings applied individually. Key findings are as follows:
\begin{itemize}

  \item The transient response of storm intensity to any combination of surface drying and roughening is well-captured as the product of the response to each forcing individually  (Eq.\eqref{vm_deconstruct}). That is, the time-dependent intensity evolution in response to a land-like surface can be understood and predicted via deconstructed physical processes caused by individual surface roughening and drying. Surface roughening imposes a strong and rapid initial response and hence dominates decay within the first few hours regardless of the magnitude of drying. 
  
  \item The equilibrium response of storm intensity to simultaneous surface drying and roughening is well-predicted by traditional potential intensity theory (Eq.\eqref{predictedVp}).
  
  \item The transient response of storm intensity to drying and roughening can be predicted by the intensification theory of Emanuel (2012), which has been generalized to apply to weakening in this work (Eq.\eqref{Eq_E12_decay1}). This theory predicts an intensity decay to a final, weaker equilibrium that can be estimated by Eq.\eqref{predictedVp}. The intensity prediction also depends on the boundary layer depth scale $h$, whose best-fit values are comparable to the value used in E12. Systematic trends of $h$ to surface forcings do not clearly match commonly-defined TC boundary layer height. 
  
  \item An additional modification is required to model the near-surface (50m) response specifically for surface roughening, which induces a rapid initial decay for near-surface intensity during the first 10 minutes. The magnitude of this initial rapid response increases with enhanced roughening and can be modeled analytically as a pure frictional spin-down (Eq.\eqref{Eq_KEbudget2}-\eqref{Eq_E12_decay2}).
 
  \item The above findings about the transient and equilibrium responses can be applied together to generate a theoretical prediction for the time-dependent intensity response to any combination of simultaneous surface drying and roughening (Eq.\eqref{E12_and_E86combine}). This prediction compares reasonably well against simulation experiments with both surface forcings.
  
  \item In the special case where the final equilibrium intensity is taken to be zero, the E12 solution reduces to a simpler analytic form that depends only on initial intensity and boundary layer depth (Eq.\eqref{Eq_E12_vp0}). This solution is found to compare well against experiments with surface fluxes turned off for a range of magnitudes of surface roughening. It also compares well with the prevailing empirical model for landfall decay (Eq.\eqref{Eq_empirical}) across a range of initial intensities. 
\end{itemize}

Although existing intensity theories are formulated for the tropical cyclone over the ocean, the above findings suggest that those underlying physics may also be valid in the post-landfall storm evolution. Note that we have not systematically tested the underlying assumptions of the theory but have focused on testing the performance of theories for predicting the response to idealized landfalls. The principal result is that for an idealized landfall, one can generate a reasonable prediction for the time-dependent intensity evolution if the inland surface properties along the TC track are known. 

Landfall in the real world is certainly much more complicated. The real world has substantial horizontal variability in surface properties compared to the simplified idealized landfalls where only the surface roughness and wetness beneath the storm are instantaneously and uniformly modified. Additional environmental variability during the transition, including heterogeneity in surface temperature and moisture, environmental stratification, topography, land-atmosphere feedbacks, vertical wind shear, and translation speed, is excluded in these idealized simulations. Therefore, a theoretical prediction for the first-order intensity response to major post-landfall surface forcings in an idealized setting provides a foundation for understanding TC landfall in nature. In this vein, our results suggest that the TC landfall process could plausibly be deconstructed into transient responses to individual surface and/or environmental forcings as encoded in our existing theories. One implication of this work is the potential to predict how post-landfall intensity decay may change in a changing climate if we know how each surface forcing will change in the future \citep{Zeng2020}. Theoretical solutions presented in this work could also benefit current risk models for hazard prediction. 

In terms of theory, future work may seek to test the theory against simulations in three-dimensional and/or coupled models that include additional complexities. That said, several questions pertinent to axisymmetric geometry remain open here: do changes in surface sensible heat fluxes significantly alter the response to surface drying? How might changes in $C_k$, whose variation after landfall is not known, alter the results? How should one optimally define TC boundary layer height for the convective eyewall region where boundary layer air rises rapidly into updrafts, both in general and in the context of the transition from ocean to land? How best can this be used to approximate the boundary layer depth scale $h$ in the theoretical solution?

Notably, the solution for pure frictional spin-down (Eq.\eqref{Eq_KEbudget2}) and the E12 solution for zero final intensity (Eq.\eqref{Eq_E12_vp0}) have an identical mathematical form, with the lone difference being trading the parameter $C_k$ for $2C_d$. These are simply the exchange coefficients for the dominant kinetic energy source (enthalpy fluxes) and sink (frictional dissipation) for the TC, respectively. Physically, we interpreted these two solutions as found in our work as a transition from a rapid response governed by pure frictional spin-down to a response governed by the reintroduction of the counterbalancing thermodynamic source of energy for the tropical cyclone as encoded in Emanuel (2012) theory (and similarly in traditional time-dependent Carnot-based theory). More generally, though, why should the large difference in the underlying physics of these two regimes manifest itself mathematically as a simple switch in exchange coefficients? This is curious.
 
Finally, future work may seek to test these theoretical predictions against observations accounting for variations in surface properties. Here we showed that our physically-based model appears at least broadly consistent with the prevailing empirical exponential decay model, suggesting that our model may provide an avenue for explaining variability in decay rates both spatially and temporally, including across climate states. For example, theory may be useful to understand how surface properties facilitate those rare TCs that do not weaken after the landfall \citep{Evans2011, Andersen+Shepherd2013}. This would help us link physical understanding to real-world landfalls, which is important for improving the modeling of inland hazards.


\acknowledgments
The authors thank for all conversations and advice from Frank Marks, Jun Zhang, Xiaomin Chen on the hurricane boundary layer. The authors were supported by NSF grants 1826161 and 1945113. We also thank for all feedbacks and conversations related to this research during 101\textsuperscript{th} AGU and AMS annual fall meetings.


\appendix%
The E12-based decay solution is derived from Eq.17 of Emanuel (2012), given by
\begin{equation}\label{appendix1}
\frac{\partial v_m}{\partial \tau} = \frac{C_k}{2h}({v_{f}}^2-{v_m}^2)
\end{equation}
where $v_m$ is the initial intensity (maximum tangential wind speed) and $v_{f}$ is defined as the theoretical steady-state maximum intensity. However, $v_{f}$ need not be larger than the current intensity but rather may be generalized to any final quasi-steady intensity, larger or smaller. Integrating Eq.A1 yields:
\begin{equation}\label{appendix2-1}
\int \frac{1}{{v_{f}}^2-{v_m}^2} \,d{v_m} = \int \frac{C_k}{2h} \,d\tau
\end{equation}
\begin{equation}\label{appendix2-2}
\frac{\ln\left|v_m+v_{f}\right|-\ln{\left|v_m-v_{f}\right|}}{2v_{f}}  = \frac{C_k}{2h}\tau+C
\end{equation}
\begin{equation}\label{appendix2-3}
\frac{\left|v_m+v_{f}\right|}{\left|v_m-v_{f}\right|}  = e^{(2\frac{C_kv_{f}}{2h}\tau+C)}
\end{equation}
\cite{Ramsay2020} showed that for intensification where $v_m<v_{f}$, the solution is
\begin{equation}\label{appendix2-4}
\begin{split}
v_{m(th)}(\tau)&=v_{f}\left(\frac{e^{(2\frac{C_kv_{f}}{2h}\tau+\coth^{-1}(\frac{v_{f}}{v_m(0)}))}-1}{e^{(2\frac{C_kv_{f}}{2h}\tau+\coth^{-1}(\frac{v_{f}}{v_m(0)}))}+1}\right)\\
 & = v_{f} \tanh{(\frac{C_k v_{f}}{2h}\tau+\coth^{-1}(\frac{v_{f}}{v_m(0)}))}
\end{split}
\end{equation}
Eq.A5 reduces to Eq.19 of Emanuel (2012) when $v_m(\tau=0)=0$, for which $\coth^{-1}(\frac{v_{f}}{0}) = 0$. 

Alternatively, for decay where $v_{f}<v_m$, the solution is
\begin{equation}\label{appendix2-5}
\begin{split}
v_{m(th)}(\tau)&=v_{f}\left(\frac{e^{(2\frac{C_kv_{f}}{2h}\tau+\tanh^{-1}(\frac{v_{f}}{v_m(0)}))}+1}{e^{(2\frac{C_kv_{f}}{2h}\tau+\tanh^{-1}(\frac{v_{f}}{v_m(0)}))}-1}\right)\\
 & =v_{f} \coth{(\frac{C_k v_{f}}{2h}\tau+\tanh^{-1}(\frac{v_{f}}{v_m(0)}))}
\end{split}
\end{equation}
Eq.A6 may be normalized by $v_m(0)$ to define the intensity response relative to the initial intensity.

 
 \bibliographystyle{ametsoc2014}
 \bibliography{references}

\end{document}